%% file: main.tex
\documentclass[final,5p,times,twocolumn]{elsarticle}

\usepackage{subcaption}
\usepackage{amsmath} 
\usepackage{amssymb}
\usepackage{lipsum}

\usepackage{graphicx}
\usepackage{dcolumn}
\usepackage{bm}

\DeclareOldFontCommand{\rm}{\normalfont\rmfamily}{\mathrm}
\DeclareOldFontCommand{\sf}{\normalfont\sffamily}{\mathsf}
\DeclareOldFontCommand{\tt}{\normalfont\ttfamily}{\mathtt}
\DeclareOldFontCommand{\bf}{\normalfont\bfseries}{\mathbf}

\usepackage{enumitem}
\newlist{alphalist}{enumerate}{1}
\setlist[alphalist,1]{label=\textbf{(\alph*)}}
\usepackage{longtable}
\usepackage{lipsum} 
\usepackage{hyperref}
\usepackage{amsmath}

\hypersetup{
    colorlinks=true,
    linkcolor=green,
    citecolor=blue,
    urlcolor=blue,
}

\hypersetup{
        pdftitle={Evidence of Relationships Among Fundamental Constants of the Standard Model},
        pdfsubject={Evidence of Relationships Among Fundamental Constants of the Standard Model},
        pdfauthor={S. V. Chekanov and H. Kjellerstrand},
        pdfkeywords={Standard Model constants, quarks, vector bosons, genetic programming, artificial intelligence}
    }
    
\journal{HEP-ANL-198667}

\begin{document}

\begin{frontmatter}


\title{Evidence of Relationships Among Fundamental Constants of the Standard Model}

\author[a]{S. V. Chekanov}
\affiliation[a]{organization={HEP Division, Argonne National Laboratory},
            addressline={9700 S.Cass},
            city={Lemont},
            postcode={60516},
            state={IL},
            country={USA}}
\ead{chekanov@anl.gov}

\author[b]{H. Kjellerstrand}
\affiliation[b]{organization={hakan.org},
            addressline={Sodra Forstadsgatan 40b},
            city={Malmo},
            postcode={21143},
            country={Sweden}}
\ead{hkjellerstrand@acm.org}



\begin{abstract}
This paper presents an approach to reducing the number of fundamental parameters in the Standard Model (SM) using genetic programming, a machine learning technique based on evolutionary algorithms. We outline the core principles of our method and identify the simplest analytic relationships among SM parameters.
Our results suggest that the SM parameters associated with quark and boson masses are not randomly distributed, but instead follow a hierarchical structure within a high-dimensional functional space. The found analytic solution depends on only two input parameters, representing the simplest mathematical model that could provide a foundation for developing a future theoretical framework to address the SM.
\end{abstract}

\begin{keyword}
Standard Model constants, quarks, vector bosons, genetic programming, artificial intelligence

\PACS{11.15.-q \sep 12.15.-y  \sep 12.38.-t \sep 14.65.-q \sep 14.70.-e \sep 06.20.Jr \sep 06.20.Jr}

\end{keyword}

\end{frontmatter}




\input{chapIntro} 

\input{chapTechnical} 

\section*{Acknowledgments}
The submitted manuscript has been created by UChicago Argonne, LLC, Operator of Argonne National Laboratory (“Argonne”). Argonne, a U.S. 
Department of Energy Office of Science laboratory, is operated under Contract No. DE-AC02-06CH11357.  Argonne National Laboratory’s work was funded by the U.S. Department of Energy, Office of High Energy Physics under contract DE-AC02-06CH11357. 

\bibliographystyle{elsarticle-num}
\bibliography{references}





\end{document}

%% file: chapIntro.tex
\section{Introduction}
\label{sec:intro}

The Standard Model (SM) is considered a model rather than a complete theory, as it relies on more than 20 fundamental parameters that must be determined experimentally \cite{ParticleDataGroup:2024cfk}. This limits its potential
for observing new phenomena in particle collisions. Attempts for identification of relationships among the SM constants have a long history \cite{WBook, Nielsen:1994ab, Froggatt__2003}, but no convincing evidence has been found to support the existence of such relationships. For example, analytic connections between quark masses are not part of the SM itself. These constants are regarded as free parameters, with no obvious underlying connections.

When a theory is missing to describe apparently unconnected values, the first step is to identify the simplest analytic relationships  between such variables, before constructing a theory based on specific physics principles. In the case of SM parameters, this task is complex as it involves many parameters, some of which carry physical units. The first step in this physics program was undertaken in \cite{Chekanov:2025wzw} using symbolic regression and genetic programming (GP) - a type of evolutionary algorithm inspired by natural selection and rooted in the broader field of artificial intelligence (AI).  The symbolic regression evolves analytic expressions directly from data \cite{koza92}. By representing mathematical expressions as trees and applying evolutionary algorithms, GP explores a wide solution space unconstrained by predefined model structures, enabling the discovery of complex, previously hidden relationships. 

This study relies on the analytic relations created by GP and reported in \cite{Chekanov:2025wzw}. The datasets derived from dimensionless SM constants are well-suited for mathematical reasoning by Large Language Models (LLMs) and other AI methods,  but they are too complex for the traditional analysis of symbolic expressions employed here.
Therefore, our focus is on analytical expressions derived from constants with physical units. This choice significantly limits the number of possible solutions.

The aim of this paper is to identify the simplest analytic solutions that align with the general expectations of the SM, and to eliminate spurious relations that arise due to numerical noise. For this work, the GP approach was enhanced to incorporate dimensional analysis — a capability that was not available for the original publication \cite{Chekanov:2025wzw}. This technical advancement enables the identification of relevant analytic solutions while ensuring the verification of physical units and the preservation of dimensional consistency. 

This paper reports on analytic relations between the fundamental constants of the SM that can be regarded the simplest mathematical model. They  may serve as a basis for a future theory that will describe  SM fundamental constants with a minimal number of free parameters.

%% file: chapTechnical.tex
\section{Search strategy}
\label{search}

Table\ref{tab:table1} lists the fundamental SM parameters commonly used in particle physics, as reported by the PDG \cite{ParticleDataGroup:2024cfk}. These constants, including those with physical units (e.g., in MeV), are used as inputs for GP. Unlike Ref.~\cite{Chekanov:2025wzw}, the masses are not rescaled by the $\rho$-meson mass, as the limitation due to dimensional analysis has been overcome in the present work.

\begin{table}[ht]
  \begin{center}
    \caption{Physics fundamental constants from the PDG  \cite{ParticleDataGroup:2024cfk} used for the GP algorithm. 
     We require that the uncertainties for $m_e$, $\pi$ and $\alpha^{-1}$ are at most  a factor 100 higher than for the Higgs boson mass. 
     Making them more precise should not contribute to the results.  
    The table presents both the absolute uncertainties ($\pm \varepsilon$) and the relative uncertainties $(\pm \varepsilon^{rel}$, in percent). 
    When the PDG reports an asymmetric uncertainty, the largest value is used.
    }
    \label{tab:table1}
    \begin{tabular}{l|r|r|r|r} 
      \hline
      \textbf{Constant} & \textbf{Name} & \textbf{Value} & $\pm \varepsilon$ & $\pm \varepsilon^{rel}$ (\%)\\
      \hline
PI & $\pi$ & 3.14159 & 1e-05 & 0.0003\\
Fine-struct. (inv) & $\alpha^{-1}$ & 137.036 & 0.001 & 0.0007\\
$\alpha_s$ at $Z^0$ & $\alpha_S$ & 0.1180 & 0.0009 & 0.7627\\
\hline
  CKM constants   &    &  no units  & $\pm \varepsilon$ & $\pm \varepsilon^{rel}$ (\%)\\
\hline
12-mix angle & $\theta_{12}$ & 0.22501 & 0.00068 & 0.3022\\
23-mix angle & $\theta_{23}$ & 0.04183 & 0.00079 & 1.8886\\
13-mix angle & $\theta_{13}$ & 0.003732 & $9\times 10^{-5}$ & 2.4116\\
CP-viol. phase & $\delta$ & 1.147 & 0.026 & 2.2668\\
\hline
  Mass   &   & MeV & $\pm \varepsilon$ & $\pm \varepsilon^{rel}$ (\%)\\
\hline
electron mass & $m_e$ & 0.510998 & $10^{-6}$ & 0.0002\\
muon mass & $m_{\mu}$ & 105.658 & 0.001 & 0.0009\\
$\tau$ mass & $m_{\tau}$ & 1776.93 & 0.09 & 0.0051\\
$u$-quark mass & $m_u$ & 2.16 & 0.07 & 3.2407\\
$d$-quark mass & $m_d$ & 4.70 & 0.07 & 1.4894\\
$s$-quark mass & $m_s$ & 93.5 & 0.8 & 0.8556\\
$c$-quark mass & $m_c$ & 1273.0 & 4.6 & 0.3614\\
$b$-quark mass & $m_b$ & 4183 & 7 & 0.1673\\
$t$-quark mass & $m_t$ & 172560 & 310 & 0.1796\\
$Z$-boson mass & $m_Z$ & 91188.0 & 2.0 & 0.0022\\
$W$-boson mass & $m_W$ & 80369.2 & 13.3 & 0.0165\\
$H$-boson mass & $m_H$ & 125200 & 110 & 0.0879\\
\hline
    \end{tabular}
  \end{center}
\end{table}

A well-known feature of the SM is that the masses of fermions, as well as those of the $Z$ and $W$ bosons, approach zero when the Higgs boson mass is set to zero. This behavior arises due to their dependence on the Higgs vacuum expectation value, mediated through Yukawa couplings, which are free parameters \cite{PhysRevLett.19.1264}.
This imposes a certain structure for the possible relations between the SM parameters.

Based on the discussion above, we adopt the following strategy to identify the simplest relations between the SM constants:

\begin{enumerate}
         
    \item \textbf{Pass dimensional analysis:} 
     All solutions must pass dimensional analysis. Since our target is particle masses, the expressions on the right-hand side should have units of MeV. 
    \item \textbf{Behavior in the Limit $m_H \to 0$:} 
    All solutions for particle masses should approach zero values as $m_H \to 0$. In particular, we will focus on solutions in which a given mass is proportional to the Higgs mass.
    \item \textbf{Simplicity criterion:} 
    Among all possible solutions, we choose the ones that are the simplest, i.e. with the smallest rank defined in \cite{Chekanov:2025wzw}.  
    It is reasonable to assume that any theory capable of unifying all these parameters, if it exists, should be formulated as simply as possible, employing the fewest parameters and the least mathematical complexity. If there are several solutions with similar analytic ranks, we prefer the one that uses a constant already encountered before, thus avoiding the introduction of a new parameter, even if the overall analytic rank may slightly increase.
    \item \textbf{Refinement by precision reduction:} 
    If no acceptable solution is found for a particular mass, we reduce the precision required for that mass and repeat the GP search. 
\end{enumerate}

Let us elaborate on the last point. This step introduces a practical level of precision for candidate solutions. However, it is important to note that high numerical precision
does not necessarily translate to theoretical significance. For example, although the number $\pi$ can be calculated at an extraordinary number of decimal places, it is still just an irrational number, and the usage of all its digits is not essential for a physical theory, since digits beyond a certain level of precision have no observable effect on physical processes. It has already been observed \cite{Chekanov:2025wzw}  that the current precision for the electron mass shown in Table~\ref{tab:table1} does not lead to the GP solutions that satisfy (1) - (3), thus its precision should be reduced.

Finally, it is important to note that if we aim to express precisely known masses analytically in terms of parameters that are known with a lower precision, there are fundamental limitations. The technical challenge of deriving high-precision outputs from low-precision inputs cannot be fully overcome, and this must be kept in mind when evaluating candidate solutions.

Previously, the GP algorithm identified several simple analytic relations \cite{Chekanov:2025wzw} connecting quark and the $Z/W$ boson masses to the Higgs mass. In the {\sc Picat} implementation \cite{koza92}, a limitation of the GP required converting masses to dimensionless values by rescaling with the $\rho$ meson mass, which served as an auxiliary parameter.

In this work, we address this limitation by automatically performing dimensional analysis. Our approach, referred to as the ``GP method'', is based on the analytic snippets from symbolic regression discussed in \cite{Chekanov:2025wzw}, followed by post-processing steps to enforce the conditions 1. - 4. In the search for possible solutions, expressions containing integer constants greater than 10 were disregarded, as well as those in which one mass is related to another mass with the appearance of a single rational number serving as a coefficient of proportionality, without any other physical constants from Table~\ref{tab:table1}. The reason for this is a large number of rational numbers for multiplication to derive other masses, leading to a large number of spurious expressions.

\section{Results}
\label{sec:result}

The GP method yields the following simplest solution connecting the SM masses with the Higgs mass:

\begin{align}
m_u &= m_H \> / (3\> (\alpha^{-1})^2),  &r=17, \label{m1_u}  \\
m_d &= m_H \> \theta_{13}^{3/2} / 6,  &r=22, \label{m1_d} \\
m_s &= m_H \>  \theta_{13} / 5,  &r=12, \label{m1_s} \\
m_c &= m_H \> \theta_{13}^{1/2} / 6, &r=17, \label{m1_c}\\
m_b &= m_H \> 9\, \theta_{13}\,(1 - \theta_{13}), &r=21, \label{m1_b}\\
m_t &= m_H \>  (\delta + \delta / 5), &r=16, \\
m_W &= m_H / \arctan( (10 - \delta)^2), &r=22, \label{m1_W}\\
m_Z &= m_W \> \delta \> \cos(\delta - 1), &r=22. \label{m1_Z}
\end{align}
The analytic rank, denoted by the letter $r$ and defined in \cite{Chekanov:2025wzw}, is indicated for each relation.  The total rank of this system of equations is 149. It is calculated as the sum of all individual ranks.
For the found solution, 8 masses are connected via $m_H$ and 3 other parameters: 
$\alpha^{-1}$, $\delta$ and $\theta_{13}$ from Table~\ref{tab:table1}.
The mass of $m_Z$ agrees with the PDG value when using
$\delta = 1.1472$, i.e. within its experimental uncertainty.
Note that the connection between $m_Z$ with $m_H$ was not found. 

We have also found a second solution, but its overall rank was much higher, $154$. Another, a third  solution, had a lower total rank $137$, but the price to pay was the introduction of a new parameter, 
$\alpha_S$. 

The obtained  relations are the result of the GP method, thus all masses agree with Table~\ref{tab:table1} within the PDG uncertainties. 
It is interesting that when expressing the masses via the Higgs mass, the parameter $\theta_{13}$ becomes a dominant constant, which may suggest certain dynamical effect.


We also identified relations in which each mass was proportional to $m_t$, $m_W$, or $m_Z$. The calculated total analytic ranks were 142, 142, and 141, respectively. We did not find a solution with fewer than three functional arguments. Thus, such solutions are consistent in terms of analytic complexity with Eqs.~(\ref{m1_u})–(\ref{m1_Z}). 

The individual relations in which each mass is proportional to $m_e$ have ranks in the range 18–23. However, we could not obtain a complete system of relations for all 8 masses, since no solution has been found in which $m_W$ is proportional to $m_e$.

\section{Alternative solution}
\label{sec:alt}
In general, we expect a large number of possible solutions, which are tedious to examine without dedicated AI analyses. 
In this study, we have chosen to focus on the simplest case, where the masses exhibit a hierarchical dependence on one another and are not directly related to the Higgs mass.
In this case, one can treat $m_e$ as the argument that defines the mass of the light quarks. Although such a relation is unknown in the SM, it is proposed in the Grand Unified Theories (see \cite{Raby:2006sk} for a review).

The GP identifies the relation $m_u = m_e \> (\pi +1 )$ as having the smallest analytic rank of 6, while preserving the dimensional condition.  The above solution satisfies our requirement of $m_u=0$ when $m_H \to 0$, since the electron mass in the SM is proportional to the Higgs mass through the Yukawa coupling.

In the alternative model, all particle masses were arranged in increasing order, starting with quarks (from the up quark to the top quark) and then vector bosons ($W$ and $Z$). In the SM, the top quark decays almost exclusively via a $W$ boson (and a $b$ quark), since the weak interaction couples quarks of different generations through the CKM matrix. Therefore, it is natural to expect a connection between the $m_t$ and $m_W$ masses. The top quark and the $Z$ boson are also linked in the SM, as the $Z$ boson couples directly to a top–antitop pair. Furthermore, the masses of $W$ and $Z$ bosons in the SM are related at tree level via $m_W = m_Z \cos(A)$, where $A$ is the Weinberg angle \cite{PhysRevLett.19.1264}. Thus, the GP approach encounters an ambiguity regarding the relationship between $m_W$, $m_Z$, and $m_t$.

Our analysis of the GP results finds the simplest solution that links each mass to the preceding one:

\begin{align}
m_u &=m_e\> (\pi +1 ),   &r=10, \label{m_u} \\
m_d &= m_u \>  (\delta +1),   &r=10, \label{m_d} \\
m_s &= m_d \>  (6 \pi  + 1),  &r=16, \label{m_s} \\
m_c &= m_s \>  (4 \pi + 1), &r = 16, \label{m_c} \\
m_b &=  m_c \> ( \pi + 1/7), &r=17, \label{m_b}  \\
m_t &= m_b \> \pi (\pi +10), &r=16, \label{m_t}  \\
m_W &= m_t \> / (\delta + 1),  &r=11,  \label{m_W} \\
m_Z &= m_W \> \delta \> \cos(\delta - 1), &r=22.  \label{m_Z} 
\end{align}
The analytic rank of each relation is indicated. The sum of these ranks is 118.
For these 8 relations, the fundamental masses are expressed using only $m_e$ and $\delta$. 
Setting $m_e = 0$ forces all these masses to vanish, in accordance with our main search criterion - that the Higgs mass determines the masses of all leptons, quarks, and vector bosons.
To continue the same logic when an equation contains a lower-mass particle, the GP method reports the following relations: $m_H = m_W\>  (6 / (5 - \delta))$ and $m_t = m_H\> (\pi + 1) / 3$.

The alternative solution Eqs.~(\ref{m_u})-(\ref{m_Z})  differs drastically from Eqs.~(\ref{m1_u})–(\ref{m1_Z}). It suggests that the equations become analytically simpler when the masses are related through a hierarchical chain, from low to high mass, rather than when each mass is directly proportional to the Higgs mass (or to the $m_t$, $m_W$ or $m_Z$ mass).
In addition, this alternative solution has only two measured parameters ($\delta$ and $m_e$), with a prominent appearance of $\pi$ and $(\delta \pm 1)$. The latter may suggest a mathematical pattern, potentially indicative of underlying physical mechanism or some hidden symmetry.

According to the definitions of the GP algorithm, the equations Eqs.~(\ref{m_u})-(\ref{m_Z}) hold true considering each argument to be taken from Table~\ref{tab:table1}.  However, when considering them as a self-contained system of equations, they
exhibit some discrepancies with the 
experimental masses.
The input values $m_e =  0.510998(1)$  and $\delta = 1.147(26)$ for Eqs.~(\ref{m_u})–(\ref{m_Z})
lead to some tensions with Table~\ref{tab:table1} when each mass is evaluated using the preceding expressions. This is because small shifts within the uncertainty of each mass pile up in the hierarchy of relations.
For example, the accumulation of the shifts in the chain of the values
produces $m_Z=87827$ MeV, which is far away from the nominal mass of the $Z$ boson.
Thus, Eqs.~(\ref{m_u})–(\ref{m_Z}), with a total rank of 118, can be considered an ideal simplest mathematical model obtainable by the GP method.

The solution for the above problem can be the following. Since we already know how the structure of the relations should look, one can re-run the GP method a second time, when using the masses as input for each relation in 
the chain Eqs.~(\ref{m_u})-(\ref{m_Z}), and vary the input parameter $\delta$ within its allowed experimental range. The obtained solution is:

\begin{align}
m_u &=m_e\> 4 \sqrt{\delta},   &r=17, \label{m2_u} \\
m_d &= m_u \>  (\delta +1),   &r=10, \label{m2_d} \\
m_s &= m_d \>  9\, \tan(\delta),  &r=18, \label{m2_s} \\
m_c &= m_s \>  (4 \pi + 1), &r = 16, \label{m2_c} \\
m_b &=  m_c \> ( \pi + 1/7), &r=16, \label{m2_b}  \\
m_t &= m_b \> \pi (\pi +10), &r=16, \label{m2_t}  \\
m_W &= m_t \> / (\delta + 1),  &r=11,  \label{m2_W} \\
m_Z &= m_W \> \delta \> \cos(\delta - 1), &r=22,  \label{m2_Z} 
\end{align}
assuming $\delta=1.1471$, i.e. within the allowed uncertainty. 
This set of relations has the required property: Each derived mass is within the experimental uncertainty, and uses the mass calculated from the preceding expression. The total analytic rank is 126. It is somewhat larger than for Eqs.~(\ref{m_u})-(\ref{m_Z}), but it is still substantially smaller than for Eqs.~(\ref{m1_u})-(\ref{m1_Z}), or any solution where $m_t$, $m_Z$ or $m_W$ are used as the multiplicative factor.

Although the above relations precisely reproduce the measured masses from Table~\ref{tab:table1}, one can obtain simpler relations
for the light-quarks masses, by sacrificing their accuracy.
It should be mentioned that the masses of light quarks are the least precisely measured. They are evaluated indirectly in the specific subtraction scheme ($\overline{MS}$), and are somewhat contentious \cite{ParticleDataGroup:2024cfk}.   Often, the elegance of a mathematical description is more important than an exact match to experimental values.
However, since we do not have a guiding physics principle for simplifying these relations, we will refrain from proposing relations for light quarks
with lower analytic ranks.

\section{Closure test}
\label{sec:closure}
It is important to check how well the analytic rank reflects the underlying connection between masses — i.e., whether the small overall analytic rank  observed for Eq.~(\ref{m_u}) - (\ref{m_Z}) is a reflection of the existing functional structure. 
We used the following simple benchmark model with six constants defined by the connected equations:
\begin{align} 
d &=c \> \sqrt{a},  \quad 
e =d \> (2 + \pi),  \quad
f =e \> b - e. 
\label{eq10}
\end{align}
Here,  $d$, $e$ and $f$ are  determined by $a$, $b$ and $c$ constants.   The total analytic rank of this mathematical model is 32.  The input for GP was created by Eq.~(\ref{eq10}) as:
\begin{align}
a &=9.15(2),  \quad b =5.24(1),  \nonumber \\
c &=3.335(1)~\mathrm{u}, \quad d = 10.088(2)~\mathrm{u}, \nonumber \\
e &=51.868(5)~\mathrm{u}, \quad f = 219.92(1)~\mathrm{u}, \nonumber
\end{align}
where $a$, $b$ and $c$ are selected arbitrarily. $a$ and $b$ do not have physical units, while other values have the same physical unit denoted with "u". The uncertainties are given in round brackets.
Note that the uncertainties of $d$, $e$ and $f$ are not given by the equations but are set to reproduce the experimental behavior of uncertainties similar to the SM masses, i.e. in the decreasing order relative to masses. 

We ran the GP algorithm and searched for the solutions that passed the dimensional analysis and have the following characteristics: 

\begin{alphalist}

\item
Solutions where every constant with a physical unit is proportional to another constant with the unit, that is, $c$, $d$, $e$ or $f$.  For example, one solution is where all three constants $d$, $e$ and $f$ are proportional to $c$, such that they vanish when $c\rightarrow 0$.  Note that there are several combinatorial combinations of this structure. Such solutions are  similar to Eqs.~(\ref{m1_u}) - (\ref{m1_Z}).

\item
Solutions organized in hierarchical order, from lowest to highest values, but proportional to a value of the previous relation, i.e. as in Eq.~(\ref{eq10}), or similar to  Eqs.~(\ref{m_u}) - (\ref{m_Z}).

\end{alphalist}
Complete sets of equations with the behavior {\bf (a)}, with 3 relations in each,  were not found up to the analytical rank of 40 of the individual expressions, suggesting that the total analytic ranks of the ``non-hierarchical'' system  are significantly larger than for Eq.~(\ref{eq10}). Most of the separate relations, which did not form a complete solution with a set of 3 equations in each, had the analytic ranks 18 and above.

The GP method found the solution {\bf (b)}, identical to  Eq.~(\ref{eq10}), with the lowest rank 32 among all possible solutions.
We repeated this test with several values generated using Eq.~(\ref{eq10}), including uniform relative uncertainties, and for each test we could identify this system of relations.
Thus, if the underlying mathematical behavior exhibits a hierarchical structure of connected values, similar to Eq.~(\ref{eq10}), it would be observed in this exact form when analyzing the GP relations. Solutions of the type {\bf (a)} would not be preferable, as their analytic ranks are significantly higher than those of {\bf (b)}.
This test confirms that the GP method accurately reproduces the underlying mathematical relationships.

\begin{figure*}[h!] 
    \begin{minipage}[b]{0.45\textwidth} 
        \centering
        \includegraphics[width=\linewidth]{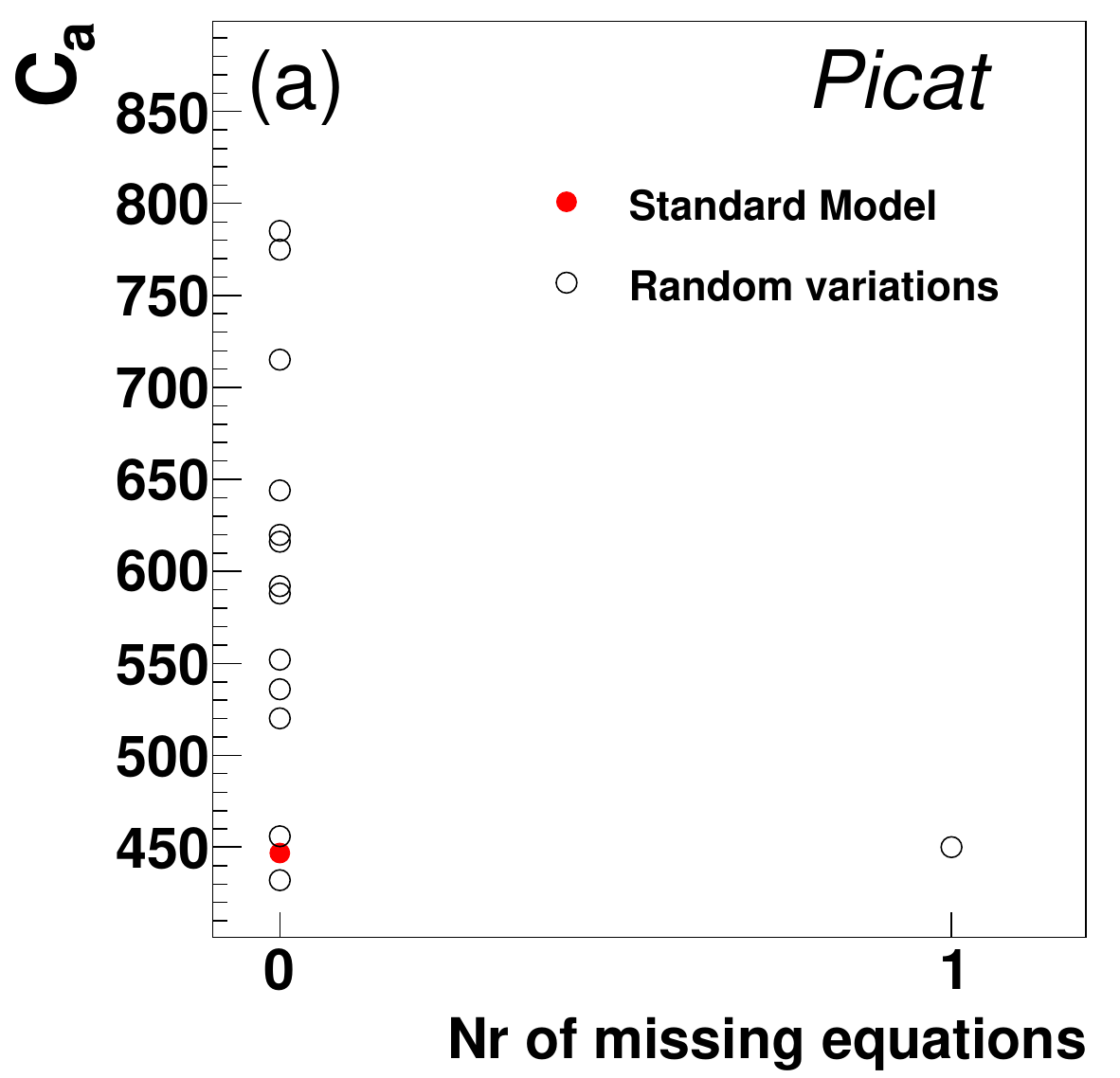}
        \subcaption{ Solution type {\bf (a)}.}
        \label{fig:imageA}
    \end{minipage}
    \hfill 
    \begin{minipage}[b]{0.45\textwidth} 
        \centering
        \includegraphics[width=\linewidth]{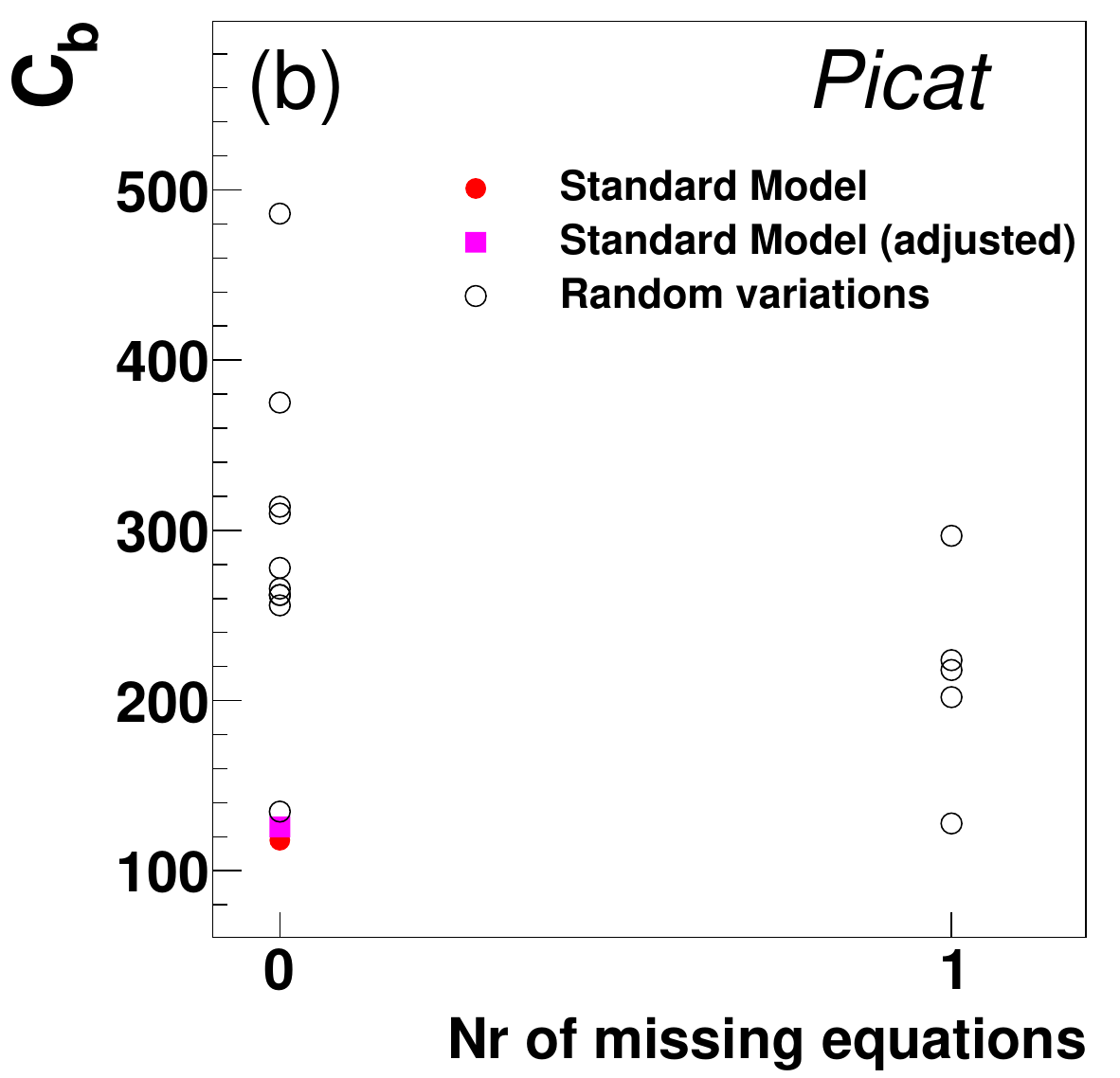}
        \subcaption{ Solution type {\bf (b)}.}
        \label{fig:imageB}
    \end{minipage}
    \caption{The summary of random variations of the SM parameters around their nominal values (open symbols). 
    The $Y$-axis shows the total complexity of the system of equations, $C$, defined as the product of the total analytic rank and the number of free parameters. 
    The filled symbols show the system of equations using SM parameters, i.e.
     Eqs.~(\ref{m1_u}) - (\ref{m1_Z}) (Fig.~\ref{fig:imageA}) and Eqs.~(\ref{m_u}) - (\ref{m_Z})  
     (Fig.~\ref{fig:imageB}). The symbols marked with ``adjusted'' indicate the solution Eqs.~(\ref{m2_u}) - (\ref{m2_Z}).
    The randomized data were processed with the help of a  {\sc Picat} code and then  the solution  type  {\bf (a)}  and {\bf (b)} were found. When some relations cannot be identified, $C$ cannot be compared on equal footing with the values of the complete equation systems.} 
    \label{fig:combined_figures}
\end{figure*}

\section{Random sampling test}
\label{sec:random}
It was also examined whether the preference for the chain structure {\bf (b)} (or  Eq.~(\ref{m_u})–(\ref{m_Z})), over the multiplicative solution {\bf (a)} is merely an artifact of the ordering of relative experimental errors in Table~\ref{tab:table1}, since these errors typically decrease with increasing SM mass.

We created 15 pseudo-experiments. Each experiment had 19 parameters with the same names as in Table~\ref{tab:table1}. The values of the  CKM constants and the masses of quarks and bosons were generated using a uniform random distribution in the range [$d_i - X_i \varepsilon_i,\>  d_i + X_i  \varepsilon_i$], where $d_i$ are the nominal SM constants, $\varepsilon_i$ are the corresponding experimental uncertainties, and $X_i$ are scaling factors chosen to ensure sufficient random smearing.
The $X_i$ was set to 20 for the CKM matrix and the $u$-quarks mass. Then $X_i$ is increased by 20 for every subsequent mass, i.e. it is set to 40 for the $d$-mass, 60 for the $s$-mass and so on, up to 180 for the Higgs mass. This approach provides significant random smearing of the values around the nominal SM constants, but prevents overlap in values, i.e. it preserves the hierarchical structure of masses, from the smallest mass for the $u$-quarks to the heaviest mass for the $t$-quarks. For every random value, the uncertainties were re-calculated to maintain the same relative uncertainties $\pm \varepsilon^{rel}$ as in Table~\ref{tab:table1}.
We limit the GP production of analytic relations only to the SM masses, producing about 400,000 snippets per random test.  Each set required 5,000 CPU hours with the Intel E5-2650 v4 processor for the {\sc Picat} program \cite{PicatWebsite}. The produced GP analytic snippets are available in \cite{GitRepo}. Then, solution types  {\bf (a)} and {\bf (b)} were automatically identified within the generated GP data.

First, we checked the analytic connection between $m_Z$ and $m_H$, which was not found for the SM parameters in Sect.~\ref{sec:result}.
However, we did find such expressions in 4 out of 15 random sets, indicating that the probability of observing such relations is not small. After this test, we focus only on the relationship between $m_Z$ and $m_W$, as in Eq.~(\ref{m1_Z}) and (\ref{m_Z}).

The summary of random variations of the SM parameters around their nominal values is presented in Fig.~\ref{fig:combined_figures}. The relations were searched up to the analytical rank 40 of the individual relations.
The $X$-axis represents the completeness of the system of relations, with 0 indicating full completeness.
The $Y$-axis shows the analytic rank of the equation system multiplied by the number of free parameters, denoted
with the symbol $C$. 
When calculating the number of free parameters, we do not include $m_H$ (for the solution {\bf (a)}) and $m_e$ (for the solution {\bf (b)}).  These two masses are used for setting all masses to zero to satisfy the SM requirement. 
The value of $C_a$ (for the solution {\bf (a)}) or $C_b$ (for the solution {\bf (b)}) serves as a measure of the complexity of the system of relations. This value, however, is  unknown\footnote{We do not exclude the existence of the solutions beyond rank 40, but such expressions are expected to be  complex, and may not be too useful for our comparison.}  for the solutions that miss one or two relations.
Such missing relations are most often associated with the $m_t$, $m_W$, and $m_Z$ masses. 

We did not attempt to use random data to reconstruct the ``connected'' system, as shown 
in Eqs.~(\ref{m2_u})–(\ref{m2_Z}), where each mass depends on the previously calculated one. The results are expected to be essentially identical to those shown in Fig.~\ref{fig:imageB}, apart from minor adjustments in the relations involving the least precisely determined light-flavor quark masses.

According to these tests, it was concluded that obtaining the complete set of 8 relations using the solution {\bf (b)} (Eqs.~(\ref{m_u}) - (\ref{m_Z}))  is more difficult, than for the solutions {\bf (a)} (Eqs.~(\ref{m1_u}) - (\ref{m1_Z})).  However, the solution {\bf (b)} is easier to find with fewer free parameters, though many such solutions do not lead to systems with 8 equations.  One test for the solution {\bf (a)} showed a better total rank, with 3 independent variables, see Fig.~\ref{fig:imageA}. 
But this random dataset has the value $C_b = 287$ in Fig.~\ref{fig:imageB}, which is far from $C_b = 118$ of the SM solution type {\bf (b)}.

To estimate the probability of observing the overall functional complexity of Eqs.~(\ref{m1_u})–(\ref{m1_Z}) and (\ref{m_u})–(\ref{m_Z}) from the 15 random tests, we calculated the distribution of $C_a + C_b$ for each test. In cases where one relation was not found, we calculated $C_{a,b}$ from the incomplete system of equations, incrementing the number of free parameters by one, treating it as a hidden variable that restores the full set of equations. The resulting distribution was symmetric, consistent with a normal distribution, with a mean of 890 and a width of 125. For comparison, the SM corresponds to a total complexity value of $C_a + C_b = 565$, which was obtained adding $146\times 3$ (for Eqs.~(\ref{m1_u})–(\ref{m1_Z})) and $118$ (for Eqs.~(\ref{m_u})–(\ref{m_Z})). The probability of obtaining a value as low as 565 from the random tests, assuming a one-sided normal distribution, is $0.46\%$.
For the most likely solution {\bf (b)} and  Fig.~\ref{fig:imageB}, the distribution of $C_b$  has a mean of 304 and a width of 76. The one-sided probability of observing $C_b\leq 118$ is 0.72\%. Thus, our tests provide evidence against the possibility of reproducing Eqs.~(\ref{m1_u})–(\ref{m1_Z}) and (\ref{m_u})–(\ref{m_Z}) in the random experiments by chance. A stronger conclusion may be possible in the future with greater CPU availability and a more autonomous evaluation of the GP expressions using AI.

In summary, we conclude that the obtained GP relationships among the SM parameters are unlikely to be numerical artifacts, but rather carry the signature of an underlying theory that unifies these fundamental constants.

\section{The lepton masses}

The lepton masses are the most difficult part of the analysis. 
The GP method could not identify connections of the electron mass
with the Higgs mass, unless the experimental uncertainty on the electron mass is drastically increased, and we have to use our 4th principle ``Refinement by precision reduction`` mentioned in Sect.~\ref{search}.
We assume that $m_e$ is the input free parameter to define the other lepton masses. At the lowest analytical rank, the mass of muons and $\tau$-leptons can be relatively well reproduced by the GP method:
\begin{align}
m_{\mu} &= m_t \left(\alpha^{-1} \sqrt{5 + \alpha^{-1}} \right)^{-1},\\
m_{\tau} &= m_H \> \left( 1/\alpha^{-1}  + 1/ (\alpha^{-1}  + 8) \right)
\end{align}
leading to the evaluated value of $105.659$~MeV and $1776.14$~MeV. Although such masses are slightly outside the modern experimental uncertainties, their values are still quite precise. The analytic rank 21 and 20 of these expressions, on average,  are higher than for the most expressions related to the quark masses.

\section{Conclusion}

We present a promising method based on genetic programming for uncovering underlying mathematical relationships directly from data. The first step follows a traditional symbolic regression approach: deriving analytical expressions from the data and generating a large set of candidate relations \cite{Chekanov:2025wzw}. The second step involves identifying simplest connecting patterns among these relations, dominated by numerical coincidences. This step leverages dimensional analysis and general SM expectations to reveal possible analytic structures within a high-dimensional functional space.

The above method was applied to the fundamental constants that are presently treated in the SM as free parameters.  
We considered two possible solutions that satisfy the key feature of the SM: when the Higgs mass $m_H$ is set to zero, due to its connection to the vacuum expectation value, all other masses also become zero.

One remarkable observation is that the SM parameters prefer
to be related in a chain of equations, i.e. each mass is connected to a particle with a lower mass, rather than to the Higgs boson mass directly.
This ``hierarchical'' structure exhibits the lowest analytic rank among all combinations examined in this paper.
It is also interesting that the dominant constant in Eq.~(\ref{m1_u}) - (\ref{m1_Z}) is $\theta_{13}$, while the alternative solution Eq.~(\ref{m_u}) - (\ref{m_Z}) is dominated by $(\delta\pm 1)$ and the constant $\pi$.
Our tests provide statistical evidence that values randomly smeared around the nominal SM constants are unlikely to reproduce the complexity of the two proposed systems of equations.

All of the above suggests that the SM parameters are not entirely random. 
Instead, they are likely interconnected within a high-dimensional functional space, pointing to an underlying dynamical mechanism or symmetry pattern that has yet to be fully understood. 

In the alternative solution, the eight masses of the SM particles  (six quarks and two vector bosons)  are expressed using only two constants: 
$\delta$ and $m_e$. This represents a substantial reduction in the number of free parameters. It should be emphasized that the found system of relations is a simplest mathematical model guided by the data from GP, rather than a physics model, since assumptions about specific symmetry principles or dynamical features were not used. We find these results sufficiently intriguing, as they demonstrate the potential in reducing the number of the SM constants to just a few. 
Whether these expressions constitute a viable theoretical framework, or will continue to hold as the precision of measured SM constants improves, remains to be studied in the future.